\documentclass[preprint]{elsarticle}
\usepackage{amsmath}
\usepackage{amssymb}
\usepackage{amsthm}
\usepackage{psfrag}
\usepackage{graphicx}
\usepackage{color}
\usepackage{upgreek}
\journal{Physics Letters B}
\begin{document}
\begin{frontmatter}
%%%%%%%%%%%%%
\title{\bf  Comment about the vanishing of the vacuum energy in the Wess-Zumino model}

\author{Andrei O. Barvinsky}
\address{Theory Department, Lebedev Physics Institute, Leninsky Prospect 53, Moscow 119991, Russia}\ead{barvin@td.lpi.ru}
\author{Alexander Yu. Kamenshchik}
\address{Dipartimento di Fisica e Astronomia, Universit\`a di Bologna and INFN, Via Irnerio 46, 40126 Bologna,
Italy\\
L.D. Landau Institute for Theoretical Physics of the Russian
Academy of Sciences, Kosygin str. 2, 119334 Moscow, Russia}\ead
{kamenshchik@bo.infn.it}
\author{Tereza  Vardanyan}
\address{Dipartimento di Fisica e Astronomia, Universit\`a di Bologna and INFN, Via Irnerio 46,  40126 Bologna,
Italy}\ead{tereza.vardanyan@bo.infn.it}

\begin{abstract}
We check the cancellation of the vacuum energy in the Wess-Zumino model at the two-loop order in the component field formalisms with and without auxiliary fields. We show that in both cases the vacuum energy is equal to zero. However, in the formalism where the auxiliary fields are excluded, the vanishing of the vacuum energy arises due to the cancellation between the potential and kinetic energies, while in the formalism with the auxiliary fields, both  terms  vanish separately.
\end{abstract}
\end{frontmatter}

\section{Introduction}

The discovery of the supersymmetry -- the symmetry between bosons and fermions in the early seventies \cite{Golfand, Volkov, Volkov1}
was one of the most impressive achievements of the modern theoretical physics. In spite of the fact that  the particles, which play the role
of superpartners of the known particles belonging to the   Standard Model, are not yet observed experimentally, the interest in supersymmetry and to its applications in different
areas of physics and mathematics is still growing and the number of original papers and reviews dedicated  to different aspects of the supersymmetry is really impressive.
One of the most studied objects in this context is the Wess-Zumino model \cite{Wess-Zumino,Wess-Zumino1} -  a simple, but very rich model, where  many basic features
of the supersymmetry were observed for the first time. Amongst these features there are cancellation of the essential part of the ultraviolet divergences \cite{Wess-Zumino} and   the  vanishing of the vacuum energy \cite{Zumino-vacuum}. The fact that the vacuum energy in exactly supersymmetric theories vanishes has a rather  general nature.
Indeed, the fundamental relation defining the supersymmetric extension of the Poincar\'e algebra is
\begin{equation}
\{Q_a,\bar{Q}_b\} = 2 \gamma^{\mu}_{ab}P_{\mu},
\label{algebra}
\end{equation}
where $Q$ is the generator of the supersymmetry transformation and $P^{\mu}$ is the energy-momentum vector.
The vacuum expectation value of the energy  $\langle 0|P^0|0\rangle$ vanishes if the supersymmetry generators annihilate the vacuum state:
\begin{equation}
Q|0\rangle = 0,\ \bar{Q}|0\rangle = 0\ \Rightarrow \langle 0|P^0|0\rangle = 0.
\label{algebra1}
\end{equation}

However, it is interesting also to see how this cancellation of the vacuum energy works at the level of the concrete fields and diagrams. Indeed, in the case of a broken supersymmetry exact cancellation of the vacuum energy, discussed above, does not work; however one can still hope that at least the ultraviolet divergences in the vacuum energy expressions are cancelled due to the fact, that the numbers of the bosons and fermions present in the models under consideration are equal.

In fact, already in fifties  Pauli~\cite{Pauli} suggested that the
vacuum (zero-point) energies of all existing fermions and bosons
compensate each other. This possibility is based on the fact that
the vacuum energy of fermions has a negative sign, whereas that of
bosons has a positive one. Later,   in a series of papers
Zeldovich~\cite{zeld,zeld1} related a finite part of the vacuum energy to the cosmological constant, however rather than eliminating the
divergences via the boson-fermion cancellation, he suggested the
Pauli-Villars regularisation of all divergences by introducing a
number  of massive regulator fields. Covariant regularisation of
all contributions then leads to finite values for both the energy
density $\varepsilon$ and (negative) pressure $p$ corresponding to a cosmological constant, i.e. related by the equation of state $p =-\varepsilon$.

Recently, this Pauli-Zeldovich cancellation mechanism for the divergences of the vacuum energy due to the balance between the contributions of bosons and fermions has attracted growing attention \cite{we,we1,we-750,Visser,we-interact}. Similar questions were also discussed in papers \cite{Sirlin, Akhmedov,Nambu2,Veltman,Dudas}.
In particular, in \cite{we-interact} several models with interactions between different particles at the lowest order of the perturbation
theory were considered. Here a certain subtlety is present. To provide the cancellation of the vacuum energy divergences it is necessary to have the balance between the fermion and the boson degrees of freedom not only on shell, but also off shell.
Let us try to be more precise in the terminology. Speaking about the on shell or physical degrees of freedom, one means the number of independent Cauchy data, which can be imposed on the field under consideration. This number is associated also with the number of different particle states.
Speaking about the off shell degrees of freedom, one means instead the number of independent field components.
For example, the Dirac spinor has four complex or eight real components. When we impose the first-order Dirac equation the number of independent real components (or Cauchy data) dwindles and becomes equal to four, which corresponds to a particle and antiparticle,  having two helicity states each. Thus, one can say that  the number of degrees of freedom of a spinor field, interpreted as the number of independent field components, doubles when it is off shell. Then the  Majorana spinor has two complex components, i.e. four degrees of freedom off shell. When we  impose the first-order Dirac equation the number of degrees of freedom,  interpreted as the number of initial data for this equation, becomes equal to two ( see e.g. (\cite{Aitchison}).

We can also add that while the difference between the number of independent Cauchy data and the number of the field components for spinors is connected with the fact that they satisfy first-order differential equations, this last fact has in turn deep group-theoretical roots. Indeed, requirement that a field belongs to a certain representation of the Lorentz group implies
a definite form of the invariant wave equation for this field. Namely, the belonging of the spinor to the $(1/2,0)\oplus(0,1/2)$ representation of the Lorentz group implies that it subject to the first-order Dirac equation (see, e.g. \cite{Naimark,Gelfand}).

As is well known in the Wess-Zumino  model one has two fermion degrees of freedom of the Majorana spinor and two boson degrees of freedom associated with the scalar and pseudoscalar fields. Off shell the number of fermion degrees of freedom becomes equal to four while the role of two additional boson fields is played by  two auxiliary fields, which are independent off shell. If we consider the non-supersymmetric models with the Pauli-Zeldovich  mechanism of cancellation of UV divergences  for the vacuum energy in the presence of interactions, then the numbers of the boson and fermion degrees of freedom should also coincide not only on shell, but off shell as well \cite{we-interact}.

Thus, it looks like the introduction of the auxiliary fields becomes unavoidable even for non-supersymmetric models if we want to make the Pauli-Zeldovich mechanism efficient. Indeed, it was explicitly shown in the seminal paper by Zumino \cite{Zumino-vacuum} that  the potential  vacuum
energy (under the potential we mean the cubic and quartic  terms in the action)
in the Wess-Zumino model vanishes at the level of two-loop Feynman diagrams. Then it is easy to show (and we shall do it in Sect. 3 of the paper) that this cancellation fails in the  formalism with the auxiliary fields excluded via the equations of motion. However, explicit calculations show that in this formalism  the kinetic vacuum energy (under the kinetic we mean all the quadratic terms in the action, including mass terms) is also non-vanishing while the sum of both vanishes. Thus, all the calculations can be coherently done also in the  formalism without the auxiliary fields. Hence, the detailed presentation and comparison of the results, obtained in different formalisms for a simple but very rich Wess-Zumino model, seems to be rather instructive. We hope that these results could be useful for more complicated models as well. Besides, while the general arguments about the difference between the numbers of the on shell and off shell degrees of freedom for spinors are well
  known, especially in the context of the supersymmetric theories (see, e.g. \cite{Aitchison, Sohnius}), the concrete analysis of the manifestation of these difference in the diagram calculations are not so elaborated, at least, up to our knowledge.
Thus, it also can be of some interest.

The paper has the following structure: in the Sect. 2 we demonstrate at the two-loop order the vanishing of the vacuum energy in the Wess-Zumino model in the presence of the auxiliary fields. In Sect. 3 we do the same
in the  formalism with the auxiliary fields excluded,  while the last section contains a discussion which briefly mentions a similar derivation in the superfield formalism along with its relation to the Pauli-Zeldovich cancellation mechanism.

\section{Wess-Zumino model in the presence of the auxiliary fields:  calculation of the vacuum  energy}

The Lagrangian density of the Wess-Zumino model is
\begin{eqnarray}
&&L = \frac12(\partial_{\mu}A)(\partial^{\mu}A)
+\frac12(\partial_{\mu}B)(\partial^{\mu}B)
+\frac{1}{2}\bar{\psi}(i\hat{\partial}-m)\psi
+\frac12F^2+\frac12G^2\nonumber\\
&&+mFA+mGB+gF(A^2-B^2)+2gGAB-g(\bar{\psi}\psi A
+i\bar{\psi}\gamma^5\psi B),
\label{Lagrange-aux}
\end{eqnarray}
where $A$ is a scalar field, $B$ is a pseudoscalar field, $\psi$ is a Majorana spinor, $F$ and $G$ are auxiliary fields. The symbol $\hat{\partial}$ means $\hat{\partial}\equiv \gamma^{\mu}\partial_{\mu}$.
The propagators or contractions of these fields in the momentum representation have the  following form  \cite{Wess-Zumino, Iliop-Zum}
\begin{eqnarray}
&&D_{AA} = D_{BB} = \frac{i}{k^2-m^2},\nonumber\\
&&D_{AF}=D_{BG}=\frac{-im}{k^2-m^2},\nonumber\\
&&D_{FF}=D_{GG}=\frac{ik^2}{k^2-m^2},\nonumber\\
&&D_{\psi\bar\psi}=\frac{i(\hat{k}+m)}{k^2-m^2}.                      \label{Feynman}
\end{eqnarray}
Let us add that in contrast to Dirac spinors, Majorana spinors also have $D_{\psi\psi}$
and $D_{\bar{\psi}\bar{\psi}}$ propagators which are related to the standard ``Dirac'' propagator by the formulae
(see e.g. \cite{Srednicki}):
\begin{eqnarray}
&&D_{\psi\psi}=D_{\psi\bar{\psi}}{\cal C}^{-1},\nonumber \\
&&D_{\bar{\psi}\bar{\psi}}={\cal C}^{-1}D_{\psi\bar{\psi}},
\label{Sred}
\end{eqnarray}
where ${\cal C}$ is the charge conjugation operator. However, in all the calculations in the Wess-Zumino model the
presence of the propagators (\ref{Sred}) is boiled down only to the appearance of some additional combinatorial factors,
which multiply the standard Dirac propagator $D_{\psi\bar{\psi}}$.

The interaction Hamiltonian density of the model (we also call it  ``potential'') includes 5 terms and  is given by the expression
\begin{equation}
H_I = -gF(A^2-B^2)-2gGAB+g(\bar{\psi}\psi A+i\bar{\psi}\gamma^5\psi B).
\label{potential}
\end{equation}
The vacuum expectation value of the potential part of the Hamiltonian (\ref{potential}) is given by the expression
    \begin{equation}
    E_{\rm pot}(x) = \frac{\int [\,d\phi\,]\,H_I(x)\,e^{iS}}{\int [\,d\phi\,]\,e^{iS}}=\frac{\langle\,H_I(x)\,e^{iS_I}\,
    \rangle}{\langle\,e^{iS_I}\,\rangle},              \label{potential1}
    \end{equation}
where $S_I$ is the interaction part of the full action
    \begin{equation}
    S_I=-\int d^4y\,H_I(y)              \label{intaction}
    \end{equation}
and the brackets $\langle...\rangle$ denote the quantum average with respect to the free theory whose action is quadratic in the quantum fields $\phi^i=A,B,F,G,\psi,\bar\psi$ and incorporates $D^{ij}(k)$ as the momentum space propagator with the components (\ref{Feynman}).

The two-loop or $g^2$-order of the expression (\ref{potential1}), obtained by expanding it in $S_I\propto g$, looks like
    \begin{equation}
    E_{\rm pot}^{\rm 2-loop} = i\langle\,H_I(x) S_I\,\rangle
    -i\langle\,H_I(x)\,\rangle\,\langle\,S_I\,\rangle,    \label{potential2}
    \end{equation}
where the second term (actually vanishing because both $H_I(x)$ and $S_I$ are odd in quantum fields) subtracts the disconnected diagrams part of $E_{\rm pot}^{\rm 2-loop}$.
The expression (\ref{potential2}) corresponds to two types of the Feynman diagrams: those with the topology of a  ``dumbbell'' and those
with the ``nut'' topology.  The dumbbell  diagrams -- two tadpoles connected by the propagator,
    \begin{equation}
    \langle\,H_I(x) S_I\,\rangle^{\rm dumbbell}\propto
    D^{jn}(x,x)\,S^{(3)}_{jni}\int d^4y\,D^{ik}(x,y)\,S^{(3)}_{klm}D^{lm}(y,y),
    \end{equation}
give the vanishing contribution (here $S^{(3)}_{klm}$ is a three-vertex corresponding to the cubic interaction term (\ref{intaction})). This follows from the fact that the tadpole diagrams in the Wess-Zumino model
are cancelled due to supersymmetry. Even though this fact is well known, let us illustrate it once again for completeness.

In view of the only nonvanishing propagator components (\ref{Feynman}) the tadpoles
    \begin{equation}
    \langle\,\phi^i\,S_I\,\rangle\propto\int d^4y\,D^{ik}(x,y)\,S^{(3)}_{klm}D^{lm}(y,y)
    =D^{ik}(0)\,S^{(3)}_{klm}\int\frac{d^4k}{(2\pi)^4}\,D^{lm}(k)
    \end{equation}
with the fermion leg, $\phi^i=\psi$ or $\phi^i=\bar\psi$, vanish identically. The same is true for the tadpoles with the pseudoscalar field $B$ leg and the auxiliary field $G$ leg. The tadpole with the scalar field leg, $\phi^i=A$, is proportional at the  one-loop order to
\begin{eqnarray}
&&\langle\,A (FA^2-FB^2+2GAB -\bar{\psi}\psi A)\,\rangle
=2D_{AA}\int D_{AF}+ D_{AF}\int D_{AA}\nonumber \\
&&\qquad\quad-D_{AF}\int D_{BB}+2D_{AA}\int D_{BG}
-(-1)D_{AA}\int {\rm Tr}\, D_{\psi\bar\psi},                    \label{tadpole}
\end{eqnarray}
where for brevity we omit the momentum space integration measure.
Taking into account the formulae (\ref{Feynman}) and the fact that
    \begin{eqnarray}
    &&\int D_{AA} = iI,\ \int D_{BB} = iI \ \int D_{AF} = -im I,\  \int D_{BG} = -imI \nonumber \\
    &&\int {\rm Tr}\, D_{\psi\bar\psi} = \int \frac{d^4k}{(2\pi)^4} \frac{{\rm Tr}\,(\hat{k}+m)}{k^2-m^2} = 4imI, \ I \equiv \frac1{(2\pi)^4}\int \frac{d^4k}{k^2-m^2},
    \label{tadpole1}
    \end{eqnarray}
we see that the right-hand side of Eq.(\ref{tadpole}) vanishes.
The tadpole with the auxiliary field $F$ leg is proportional to
 \begin{eqnarray}
 &&\langle\, F (FA^2-FB^2+2GAB-\bar{\psi}\psi A)\,\rangle=D_{FF}\int (D_{AA}-D_{BB}) \nonumber \\
 &&\qquad\quad + 2D_{AF}\int D_{AF} + 2D_{AF}\int D_{BG}-(-1)D_{AF}\int {\rm Tr}D_{\psi\bar\psi} = 0.
 \label{tadpole2}
 \end{eqnarray}

Thus, considering the expression (\ref{potential2}) we take into account only the diagrams with a nut topology, i.e. the diagrams, where  two vertices are connected by three propagators. There are two types of such diagrams, ones including fermions, and those including only boson fields.
The contribution of the diagrams of the first type to the vacuum energy is given by the expression
\begin{eqnarray}
&&E_1(x) =\int d^4y\, \big\langle \,(g\bar{\psi}\psi A)(x)\cdot (-ig\bar{\psi}\psi A)(y)\nonumber\\
 &&\qquad\qquad\qquad
 +(ig\bar{\psi}\gamma^5\psi B)(x)\cdot
 (-i(i)g\bar{\psi}\gamma^5\psi B)(y)\,\big\rangle,   \label{E_1}
\end{eqnarray}
where $\cdot$ separates the factors at the points $x$ and $y$, between which and only which the chronological contractions are taken in the ``nut" diagram. This expression reads
\begin{eqnarray}
&&E_1 =16g^2\int d^4k d^4p \frac{k\cdot (p+k)}{(k^2-m^2) (p^2-m^2) ((p+k)^2-m^2)}\nonumber\\
&&\qquad=8g^2\int d^4k d^4p \frac{(p+k)^2-p^2+k^2}{(k^2-m^2) (p^2-m^2) ((p+k)^2-m^2)}\nonumber \\
&&\qquad=8g^2\int d^4k d^4p\frac{1}{ (p^2-m^2) ((p+k)^2-m^2)}\nonumber\\
&&\qquad+8g^2m^2\int d^4k d^4p \frac{1}{(k^2-m^2) (p^2-m^2) ((p+k)^2-m^2)}\nonumber \\
&&\qquad=8g^2I^2+8g^2m^2K,
\label{nut}
\end{eqnarray}
where
\begin{equation}
K \equiv  \int d^4k d^4p \frac{1}{(k^2-m^2) (p^2-m^2) ((p+k)^2-m^2)}.
\label{nut1}
\end{equation}
The sum of the contributions of the diagrams, which include only the boson fields is
\begin{eqnarray}
&&E_2 = -ig^2\int d^4y\,\left(\big\langle\,(FA^2)(x)\dot(FA^2)(y) + (FB^2)(x)\cdot(FB^2)(y)\right.\nonumber\\
&&\qquad\qquad\left.+4(GAB)(x)\cdot(GAB)(y)-4(FB^2)(x)\cdot (GAB)(y)\,\big\rangle\right)\nonumber \\
&&\qquad=-ig^2\left(2\int D_{FF}D_{AA}D_{AA}+4\int D_{AF}D_{AF}D_{AA}+2\int D_{FF}D_{BB}D_{BB}
\right.\nonumber \\
&&\qquad\left.+4\int D_{GG}D_{AA}D_{BB}+4\int D_{AA}D_{BG}D_{BG} -8\int D_{AF}D_{BG}D_{BG}\right).
\label{nut2}
\end{eqnarray}
Using the presentation
\begin{equation}
D_{FF}=D_{GG}=i+\frac{im^2}{k^2-m^2}
\label{aux}
\end{equation}
and other formulae from Eq. (\ref{Feynman}), one can reduce the right-hand side of Eq. (\ref{nut2}) to the following sum
\begin{equation}
E_2=-8g^2I^2-8g^2m^2K.
\label{nut3}
\end{equation}
In other words, the nut diagrams, including two vertices linear in auxiliary fields and bilinear in scalar (pseudoscalar) fields
give rise to two types of diagrams with simple scalar-type propagators: ones with the topology of a nut and those with the topology of the ``eight" diagram, including one 4-vertex and two one-propagator scalar loops -- the product of two one-loop integrals. Finally, we see from Eqs. (\ref{nut}) and (\ref{nut3}) that
\begin{equation}
E_1+E_2 = 0.
\label{poten2}
\end{equation}
The corresponding diagrams (without detail) were presented in the paper by Zumino \cite{Zumino-vacuum}.

What can one say about the vacuum kinetic energy in the Wess-Zumino model?
One knows that in the absence of interactions the quantum fields can be considered as the systems of free
harmonic oscillators and that the zero energy of a boson oscillator is positive while that of a fermion oscillator is negative.
As we have already mentioned in the Introduction, the requirement of the disappearance of the ultraviolet divergences in the vacuum energy of the free fields  is reduced to the equality of the numbers of  boson and fermion degrees of freedom and to certain sum rules \cite{Pauli,zeld,zeld1,we,we1,we-750}.
In the case of the supersymmetric models all the masses inside of a supemultiplet are equal and the mass sum rules are satisfied automatically. Moreover, when the interaction is switched on the masses begin their running, but their anomalous mass dimensions are such that the running masses of different particles are still equal. In the case of the Wess-Zumino
model in the formalism with the auxiliary fields there is only one renormalization constant - the wave function renomralization, which is equal for all fields. This implies the equality of the running masses. Thus, one can say
that both the kinetic vacuum energy and the potential vacuum energy in the Wess-Zumino model with the auxiliary fields are cancelled separately.

\section{Wess-Zumino model without auxiliary fields and the vacuum energy}

The Euler-Lagrange equations for the auxiliary fields $F$ and $G$ are
\begin{eqnarray}
&&F+g(A^2-B^2)=0,\nonumber \\
&&G+2gAB=0.
\label{E-L}
\end{eqnarray}
Substituting the expression for the auxiliary fields taken from Eq. (\ref{E-L}) into the Lagrangian of the Wess-Zumino model
(\ref{Lagrange-aux}), we obtain an expression which contains only the ``physical'' fields $A, B$ and $\psi$:
\begin{eqnarray}
&&L = \frac12(\partial_{\mu}A)(\partial^{\mu}A)+\frac12(\partial_{\mu}B)(\partial^{\mu}B)+\frac{1}{2}\bar{\psi}(i\hat{\partial}-m)\psi-\frac12m^2A^2-\frac12m^2B^2\nonumber\\
&&-\frac12g^2(A^2+B^2)^2-mgA(A^2+B^2) - g(\bar{\psi}\psi A+i\bar{\psi}\gamma^5\psi B).
\label{Lagrange-phys}
\end{eqnarray}
Now the interaction Hamiltonian density is
\begin{equation}
H_I = \frac12g^2(A^2+B^2)^2+mgA(A^2+B^2) + g(\bar{\psi}\psi A+i\bar{\psi}\gamma^5\psi B).
\label{poten-phys}
\end{equation}
We can now calculate the potential vacuum energy corresponding to the potential (\ref{poten-phys}). Obviously,
the contribution of the Yukawa terms will coincide with that calculated in the preceding section and  will be given by
the formula (\ref{nut}). Then we have the  ``eight"-diagram type contribution of the quartic scalar-pseudoscalar interaction
\begin{eqnarray}
&&E_3=\frac{g^2}{2}\big\langle\, (A^2+B^2)^2\,\big\rangle
=\frac32g^2\int D_{AA}\cdot \int D_{AA}
+ \frac32g^2\int D_{BB}\cdot \int D_{BB}\nonumber \\
&&\qquad+ g^2\int D_{AA} \cdot \int D_{BB}=-4g^2 I^2.
\label{E3}
\end{eqnarray}

The contributions of the triple scalar-preudoscalar interactions  have the form of the ``nut" diagram integral
\begin{eqnarray}
&&E_4=-im^2g^2\int d^4y\,\big\langle\,A^3(x)\cdot A^3(y)+AB^2(x)\cdot AB^2(y)\,\big\rangle\nonumber \\
&&\quad=-im^2g^2(6\int D_{AA}D_{AA}D_{AA} + 2\int D_{AA}D_{BB}D_{BB}) = -8m^2g^2K.
\label{E4}
\end{eqnarray}
Thus, we see that the total contribution of the potential terms into the vacuum energy is
\begin{equation}
E_1+E_3+E_4=4g^2I^2 \neq 0.
\label{nonzero}
\end{equation}

Where was the balance between the fermion and boson contributions  lost? To answer this question let us try to calculate
explicitly the vacuum kinetic energy terms. This terms will come from the expression
\begin{eqnarray}
&&E_{\rm kin} =\frac12 \left\langle\big(\,\partial_{\mu}A\,\partial_{\mu}A+m^2A^2\big)\,
e^{iS_I}\right\rangle_c+\frac12 \left\langle\big(\,\partial_{\mu}B\,\partial_{\mu}B+m^2B^2\big)\,
e^{iS_I}\right\rangle_c\nonumber \\
&&\quad\quad+\frac{1}{2}\left\langle\,
\bar\psi\big(\,i\gamma_i\partial_i+m\big)\psi\,e^{iS_I}\right\rangle_c,
\label{E-kin}
\end{eqnarray}
where $i=1,2,3$ and
    \begin{eqnarray}
    \partial_\mu A\,\partial_\mu A
    =(\partial_0 A)^2+(\partial_i A)^2                     \nonumber
    \end{eqnarray}
of course denotes Lorentz non-invariant combination contributing to the energy density of the scalar field $A$. The subscript ``c'' means that only connected diagrams should be considered,  disconnected contributions being subtracted similarly to Eq.(\ref{potential2}).

The two-loop contribution of the quartic interaction into the kinetic energy of the scalar field is
\begin{eqnarray}
&&E_5=\int d^4y\,\left\langle
\left(\frac12\partial_{\mu}A\,\partial_{\mu}A+\frac12m^2A^2\right)(x)
\cdot\left(-\frac{i}{2}g^2(A^2+B^2)^2(y)\right)\right\rangle_c\nonumber \\
&&\qquad\qquad=-4g^2I\cdot \int d^4k\frac{k_{\mu}k_{\mu}+m^2}{(k^2-m^2)^2}.
\label{E5}
\end{eqnarray}
The corresponding contribution of the quartic interaction to the kinetic energy of the pseudoscalar field is the same.

The contribution of the triple scalar (pseudoscalar) interaction to the kinetic energy of the scalar field,  which is obtained by expanding $e^{iS_I}$ to quadratic order in $mg\int d^4y\,A(A^2+B^2)(y)$, is  the following ``nut" diagram
\begin{eqnarray}
&&E_6=-\frac12m^2g^2\int d^4y\,d^4z\,\Big\langle \left(\frac12\partial_{\mu}A\,
\partial_{\mu}A+\frac12m^2A^2\right)(x)\nonumber\\
&&\qquad\qquad\qquad\cdot\left(A^3(y) \cdot A^3(z) + AB^2(y)\cdot AB^2(z)\right)\Big\rangle_c\nonumber \\
&&\qquad=-10m^2g^2 \int d^4kd^4p\frac{k_{\mu}k_{\mu}+m^2}{(k^2-m^2)^2(p^2-m^2)((k+p)^2-m^2)}.
\label{E6}
\end{eqnarray}
The sign $\cdot$ again separates groups of factors at different spacetime points, the chronological contractions inside each group being disregarded like in (\ref{E_1}).

A similar contribution of the triple scalar interaction to the kinetic vacuum energy of the pseudscalar field is different:
\begin{eqnarray}
&&E_7=-\frac12m^2g^2\int d^4y\,d^4z\,\Big\langle \left(\frac12\partial_{\mu}B\,
\partial_{\mu}B+\frac12m^2B^2\right)(x)\cdot AB^2(y)\cdot AB^2(z) \Big\rangle_c\nonumber \\
&&\qquad=-2m^2g^2 \int d^4kd^4p\frac{k_{\mu}k_{\mu}+m^2}{(k^2-m^2)^2(p^2-m^2)((k+p)^2-m^2)}.
\label{E7}
\end{eqnarray}
The contribution of the Yukawa interaction to the kinetic vacuum energy of the scalar field is
\begin{eqnarray}
&&E_8= -\frac12g^2\int d^4y\,d^4z\,\Big\langle \left(\frac12\partial_{\mu}A\,
\partial_{\mu}A+\frac12m^2A^2\right)(x)\cdot \bar\psi\psi A(y)\cdot \bar\psi\psi A(z) \Big\rangle_c\nonumber \\
&&\qquad\qquad=4g^2\int d^4kd^4p\frac{(k_{\mu}k_{\mu}+m^2)\,(p\cdot(p+k)+m^2)}{(k^2-m^2)^2(p^2-m^2)((k+p)^2-m^2)}\nonumber \\
&&\qquad\qquad=4g^2I\cdot \int d^4k\frac{k_{\mu}k_{\mu}+m^2}{(k^2-m^2)^2}\nonumber \\
&&\qquad\qquad-2g^2\int d^4kd^4p\frac{k_{\mu}k_{\mu}+m^2}{(k^2-m^2)(p^2-m^2)((k+p)^2-m^2)}\nonumber \\
&&\qquad\qquad+6g^2m^2\int d^4kd^4p\frac{k_{\mu}k_{\mu}+m^2}{(k^2-m^2)^2(p^2-m^2)((k+p)^2-m^2)}.
\label{E8}
\end{eqnarray}
The contribution of the Yukawa interaction to the kinetic vacuum energy of the pseudoscalar  field is instead
\begin{eqnarray}
&&E_9= -\frac12g^2\int d^4y\,d^4z\,\Big\langle \left(\frac12\partial_{\mu}B\,
\partial_{\mu}B+\frac12m^2B^2\right)(x)\cdot i\bar\psi\gamma^5\psi B(y)\cdot i\bar\psi\gamma^5\psi B(z) \Big\rangle_c\nonumber \\
&&\qquad\qquad=4g^2\int d^4kd^4p\frac{(k_{\mu}k_{\mu}+m^2)\cdot(p\cdot(p+k)-m^2)}
{(k^2-m^2)^2(p^2-m^2)((k+p)^2-m^2)}\nonumber \\
&&\qquad\qquad=4g^2I\cdot \int d^4k\frac{k_{\mu}k_{\mu}+m^2}{(k^2-m^2)^2}\nonumber \\
&&\qquad\qquad-2g^2\int d^4kd^4p\frac{k_{\mu}k_{\mu}+m^2}{(k^2-m^2)(p^2-m^2)((k+p)^2-m^2)}\nonumber \\
&&\qquad\qquad-2g^2m^2\int d^4kd^4p\frac{k_{\mu}k_{\mu}+m^2}{(k^2-m^2)^2(p^2-m^2)((k+p)^2-m^2)}.
\label{E9}
\end{eqnarray}
Finally, the contribution of the Yukawa interaction with scalar and pseudoscalar fields to the kinetic vacuum energy of
the Majorana spinor is
\begin{eqnarray}
&&E_{10} = -\frac12g^2\int d^4y\,d^4z\,\Big\langle \,\frac{1}{2}\bar\psi(i\gamma_i\partial_i+m)\psi(x)\cdot\big(\bar{\psi}\psi A(y)\cdot\bar{\psi}\psi A(z)\nonumber\\
&&\qquad\qquad\qquad +i\bar{\psi}\gamma^5\psi B \cdot i\bar{\psi}\gamma^5\psi B
\big)\Big\rangle_c\nonumber \\
&&\qquad\quad= 4g^2\int d^4kd^4p\frac{{\rm Tr}\,[(\gamma_i k_i +m)(\hat{k}+m)(\hat{p}+\hat{k})(\hat{k}+m)]}{(k^2-m^2)^2(p^2-m^2)((k+p)^2-m^2)}\nonumber \\
&&\qquad\quad=8g^2\int d^4kd^4p\frac{k_ik_i+m^2}{(k^2-m^2)(p^2-m^2)((k+p)^2-m^2)}\nonumber\\
&&\qquad\quad+
8g^2\int d^4kd^4p\frac{m^2}{(k^2-m^2)(p^2-m^2)((k+p)^2-m^2)}\nonumber \\
&&\qquad\quad+16g^2\int d^4kd^4p\frac{m^2(k_ik_i+m^2)}{(k^2-m^2)^2(p^2-m^2)((k+p)^2-m^2)}.
\label{E10}
\end{eqnarray}
Now, using the equality for Lorentz non-invariant combination $k_{\mu}k_{\mu}=k_0^2+k_i^2$,
\begin{equation}
2k_ik_i - k_{\mu}k_{\mu} = k^2\equiv k_\mu k^\mu,
\label{equality}
\end{equation}
one can obtain
\begin{eqnarray}
&&E_{\rm kin} = 2E_5+E_6+E_7+E_8+E_9+E_{10}\nonumber \\
&&\qquad=-4g^2\int d^4kd^4p\frac{1}{(p^2-m^2)((k+p)^2-m^2} = -4g^2I^2.
\label{E-kin1}
\end{eqnarray}
Assembling Eqs. (\ref{E-kin1}) and (\ref{nonzero}) we see that, even though separately the kinetic and potential vacuum energies in the on-shell formalism of Wess-Zumino model are nonvanishing, their sum still equals zero as it should be.

\section{Conclusions}
In this paper we have compared the two-loop cancellation of the vacuum energy in the Wess-Zumino model in different formalisms - in the component field formalism with auxiliary fields and in the component formalism without auxiliary fields. It is also well-known that zero value of the vacuum energy directly follows without explicit calculations from the superfield formalism and is based on the general properties of the superfield Feynman graphs. Without going into details we just briefly remind the superfield mechanism responsible for that \cite{F-L,O-M,O-M1}.

This mechanism holds separately for all superfield vacuum Feynman diagrams -- any vacuum diagram is equal to zero. The point is that in the absence of external  lines, a diagram under consideration depends only on the differences of the internal Grassmann variables $\theta^\alpha$. Thus, the number of Grassmann integrations is larger than the number of independent Grassmann variables and the result equals zero because of the integration rule for $\theta^\alpha$, namely $\int d\theta^{\alpha} = 0$ (see \cite{Berezin}). Thus, when we work in terms of the superfields, relatively simple algebraic manipulations allow us easily to attain the vanishing of the total vacuum energy along with a drastic decrease of the ultraviolet divergences in the
supersymmetric models. In particular, in the Wess-Zumino model it is necessary to introduce only one counter-term
- the wavefunction renormalization constant (see e.g. the review \cite{O-M1} and the references therein).

Explicit calculations in the component formalism with auxiliary fields are still not so complicated. The calculations in the formalism where the auxiliary fields are excluded by means of the equations of motion are more cumbersome and we have presented them in some detail in the section 3 of the present paper. We have seen that in this case the vacuum energy is again equal to zero. However  its vanishing is a result of the cancelation between the non-zero contributions to the vacuum energy, representing its potential and kinetic parts. Note, that in the formalism with the explicit presence of the auxiliary fields these two terms (potential and kinetic) are equal to zero separately.

Why did we undertake these calculations in spite of the fact that the vanishing of the vacuum energy in the supersymmetric models is well established many years ago? There are two reasons for that. First, it is interesting to study on one more example an involved question concerning the role of the auxiliary fields in the supersymmetric theories. To illustrate once again this problem, we shall cite some sentences from the review by Sohnius \cite{Sohnius} (p. 94):
``There are several ways to understand why on- and off-shell representations are different and why auxiliary fields appear in supersymmetric theories. At the root of the problem lies the difference between the vector space for the off-shell representations, fields over $x^{\mu}$ and that for on-shell representations, the Cauchy data for the fields equations. In both cases the fermions = bosons  rule must hold. For different spins, however, there is a different relationship between Cauchy data and fields. While, e.g., a real scalar field $A(x)$ describes one neutral scalar particle, the Dirac field $\psi(x)$ has eight real components, but describes only the four states of a charged spin - $\frac12$ particle. Thus, in going from the fields to the states, we have lost some dimensions of our representation space, but differently so for different spins. Supersymmetric models  with their strict fermions = bosons rule must somehow wiggle out of this, and they do so by means of auxi!
 liary fields whose off-shell degrees of freedom disappear completely on-shell''.

On the other hand, in recent years there has been a growing interest in the Pauli-Zeldovich mechanism \cite{Pauli,zeld,zeld1} of the cancellation of the vacuum energy divergences in the non-supersymmetric theories with equal number of fermion and boson degrees of freedom  \cite{we,we1,we-750,Visser,we-interact}. In the corresponding models some relations between coupling constants and  masses should be imposed, but one cannot use the superfield formalism. Thus, a detailed knowledge of the mechanism of the cancellations between different contributions to the vacuum energy can be of interest in this context.

\section*{Acknowledgements}
This work was supported by the RFBR grant No.17-02-00651. We are grateful to F. Bastianelli, I.V. Mishakov, A.A. Starobinsky, O.V. Teryaev, A. Tronconi, A.A. Tseytlin and G. Venturi for useful discussions.

\end{document}